# A Non-staggered Projection Algorithm for Two-Phase Fluid-Structure Interaction Simulation Using the Phase-Field/Immersed-Boundary Method

Xiaoshuang Wang [a, b, c], Liwei Tan [b], Wenjun Ying [b], Enhao Wang [c],
Yao Xiao [a], Liangqi Zhang [a, *], Zhong Zeng [a, *]

[a] College of Aerospace Engineering, Chongqing University, Chongqing, 400044, China
[b] Shanghai-Chongqing Institute of Artificial Intelligence, Chongqing, 401332, China
[c] Institute for Ocean Engineering, Shenzhen International Graduate School, Tsinghua University, Shenzhen, 518055, Guangdong, China

* Corresponding author, Liangqi Zhang: zhangliangqi@cqu.edu.cn

* Corresponding author, Zhong Zeng: zzeng@cqu.edu.cn

**Abstract**

We present a Pressure-Oscillation-Free projection algorithm for large-density-ratio multiphase fluid-structure interaction simulations, implemented on a non-staggered Cartesian grid. The incompressible Navier-Stokes is decoupled with an improved five-step incremental pressure correction algorithm. Fluid-fluid interface is captured using the Cahn-Hilliard equation, and the surface tension model is coupled with a momentum-weighted interpolation scheme to suppress unphysical pressure oscillations, ensuring accurate evolution of multiphase interfaces. Interaction at the fluid-structure interface is obtained by implicitly solving for the feedback acceleration in the Eulerian-Lagrangian system. For validation of the present method, the comparison studies for Pressure-Oscillation-Free effect are systematically conducted using lid driving cavity and droplet deformation cases. Moreover, several challenging multiphase simulations are implemented and discussed. As a demonstrating example of fluid-structure interaction, a rising bubble bypassing an obstacle is tested.

**Keywords:** Incompressible two-phase flow; Fluid-structure interaction; Phase-Field method; Immersed boundary method; Non-staggered grid; Cartesian grid;

# 1 Introduction

The interfacial dynamics of two-phase flows plays an increasingly important role in nature and many engineering and biomedical applications. Examples include bubble trajectories, tank sloshing, cavitation dynamics, and marine risers (Wu et al., 2017; Galusinski and Vigneaux, 2008; Wang et al., 2018; Lou and Liang, 2020). A fluid-fluid interface represents not only a complex state involving tension, pressure gradients, and possibly external forces but also, with respect to material properties, a jump in density and viscosity across the interface. Due to innate nonlinearities, complex topology, and the challenge of dealing with unknown, active, and moving surfaces, multiphase flow modeling is a long-standing challenge in the community of computational fluid dynamics (CFD). An accurate description of mechanical interactions between different phases necessitates the use of well-suited models and precise numerical methods.

Two primary methods for multiphase flow description are interface tracking and interface capturing. Interface tracking, based on the Lagrangian description, labels the interface with discrete points and tracks it along the fluid flow. However, it struggles with complex topological changes(Inguva et al., 2022). Interface capturing methods, such as the volume-of-fluid (VOF) (Hirt and Nichols, 1981; Gueyffier et al., 1999; Elahi et al., 2015), level-set (LS) (Osher and Sethian, 1988; Sussman et al., 1999), and phase-field methods(Jacqmin, 1999), emphasize the evolution of a distributed field variable as opposed to the direct representation of the interface itself. The Volume-of-Fluid (VOF) method employs discontinuous volume fraction values, presenting challenges in precisely computing geometric curvature and mitigating numerical dissipation. The Level-Set (LS) method requires re-initialization for significant topological changes to maintain the integrity of the distance function, potentially resulting in the violation of mass conservation for each phase. Various reconstruction schemes (Yuan et al., 2018) or coupled interface capturing algorithms(Bourlioux, 1995) have been proposed, addressing these limitations to some extent, while simultaneously intensifying the complexity of the procedural aspects. In contrast, the phase-field method avoids complicated reconstruction or re-initialization. It represents the interface using energy-based variational formalisms, preserving the total energy balance and demonstrating advantages in the theoretical foundation, implementation simplicity, and handling moving contact lines (Jacqmin, 1999; Ding et al., 2007; Shen and Yang, 2010; Sui et al., 2014). Consequently, these advantages render the phase-field method progressively appealing.

The decoupling of pressure and velocity of the Navier–Stokes equation is a key difficulty in the simulation of incompressible flow. The algorithms to date often fall into two classes, the iterative method (SIMPLE (Patankar and Spalding, 1972), PISO (Issa, 1986), etc.) and the non-iterative method (the projection method (Guermond et al., 2006), etc.). Both of them inevitably encounter the same challenge in effectively suppressing the generation of pressure oscillations. The situation becomes more intricate when delineating the interfacial dynamics arising from the nuanced equilibrium of diverse forces and pressure gradients.(Francois et al., 2006). Especially in long-term simulations, subtle pressure perturbations may evolve into a source or sink, and drive unphysical motions of the fluid interface. Historically, many algorithms have been developed to avoid pressure oscillations. The *staggered variable arrangement* (Harlow and Welch, 1965), enforces a compact coupling between pressure and velocity by storing the velocity at the centers of the cell faces, while all scalars are evaluated and placed at the cell centers. However, the staggered arrangement cannot be used for grids with skew, stretch, and compression transformations. Moreover, the complex storage mode leads to high memory requirements, which severely hampers the scaling up of computation. An alternative form of variable positioning technique, namely the *non-staggered* or *collocated grid arrangement*, stores all the variables at the same physical location and employs only one set of control volumes. The notable method that allows robust computations on a collocated grid is the momentum-weighted interpolation (MWI), also frequently referred to as pressure-weighted interpolation or Rhie–Chow interpolation (Rhie and Chow, 1983). The MWI effectively suppresses the pressure oscillation by assessing velocities at faces through weighting coefficients derived from discretized momentum equations. Numerous studies related to this field have been conducted. For example, Majumdar (Majumdar, 1988) proposed a so-called Majumdar correction to make MWI independent of the under-relaxation parameter in an iterative algorithm. Ren et al. (Ren et al., 2007) recovered the grid scale ellipticity in the pressure field with a filtering procedure. Armfield et al. (Armfield and Street, 2000) ensured an elliptic pressure coupling by including additional terms in the pressure correction equation. Denner and van Wachem (Denner and van Wachem, 2014) successfully used the time-independent interpolation approach proposed by Pascau (Pascau, 2011) for the two-phase flow based on VOF. However, numerous investigations employing the phase-field method for simulating two-phase flow persist in utilizing the staggered grid (Huang et al., 2019; Mirjalili and Mani, 2021; Sharma et al., 2021; Xia et al., 2022), despite its intricacy and computational expenses. Consequently, the phase-field method currently lacks effective techniques to suppress pressure oscillations.

In this article, we present a Pressure-Oscillation-Free approach for two-phase flow on a *non-staggered Cartesian grid*. The grid-scale ellipticity of the pressure field is recovered and thus the pressure oscillation can be removed effectively. Compared with the classical projection method, the presented interpolation algorithm is coupled with the phase-field surface tension model and only includes an additional correction step which is easy to implement and consumes negligible cost. Furthermore, the utilization of non-staggered Cartesian grids enables the convenient combination with the surface capturing algorithms, such as the immersed boundary method(Peskin, 1972) and kernel-free boundary integral method(Ying and Henriquez, 2007), providing an effective solution for treatments of complex and moving boundaries. We note that the governing equations in this article are all discretized with the finite volume method and the convection terms are consistently treated by weighted essentially non-oscillating (WENO) scheme (Jiang and Shu, 1996). The linear systems are solved by preconditioned conjugate gradients (PCG) method or fast Fourier transform (FFT) based Poisson (Helmholtz) solver.

In Section 2, we introduce the governing equations and review a classical projection algorithm for incompressible flow. Section 3 presents the non-staggered projection algorithm for two-phase fluid-structure simulations through this study. Subsequently, Section 4 presents multiple cases to verify both the accuracy and the efficacy of the proposed algorithm in suppressing unphysical oscillations. Section 5 demonstrates the algorithm's versatility across various two-phase flow and fluid-structure interaction cases. We conclude this paper in Section 6.

## 2 Theoretical backgrounds

### 2.1 Governing equations

The Navier-Stokes equations for multiphase flows with variable density and viscosity are given as

$$\nabla \cdot \vec{U} = 0 \qquad (1)$$

$$\frac{\partial(\rho\vec{U})}{\partial t} + \nabla \cdot (\rho\vec{U} \otimes \vec{U}) = -\nabla P + \nabla \cdot \left[\mu(\nabla\vec{U} + \nabla\vec{U}^T)\right] + \vec{f}_{ext} \tag{2}$$

where $\vec{U}$ is the velocity vector, $P$ is the pressure, $\rho$ is the density, and $\mu$ is the dynamic viscosity, and $\otimes$ corresponds to the tensor product. $\vec{f}_{ext}$ is the external force (such as gravity, surface tension, etc.). The divergence-free condition in Eq. (1) states the constraint of incompressibility or volume conservation. Eq.(2) is Newton's momentum law in convective fluid.

In order to capture the interface, the Cahn-Hillard equation by phase-field method (Liu et al., 2014) is applied.

$$\frac{\partial \phi}{\partial t} + \nabla \cdot (\vec{U}\phi) = \nabla \cdot (M\nabla\psi) \tag{3}$$

$\phi$ is the phase-field function ranging continuously within [-1,1], where -1 and 1 denote respectively the two phases, while the range of (-1,1) denotes the interfacial region. $M$ is the mobility, which defines the strength of diffusivity in the interfacial region. $\psi$ is the chemical potential which is derived from the Ginzburg-Landau free energy $\mathcal{F}$ (Liu et al., 2014),

$$\mathcal{F} = \int_\Omega \left[\frac{1}{4}(\phi-1)^2(\phi+1)^2 + \frac{1}{2}\varepsilon^2|\nabla\phi|^2\right]d\Omega \tag{4}$$

where $\Omega$ is the domain considered, $\varepsilon$ determines the thickness of the interface. The first term (mixing energy) represents the energy stored in the interface (Soligo et al., 2019). The second term (ideal part of the free energy) takes a double-well form and accounts for the tendency of the system to separate into two pure fluids. Therefore, the expression of the chemical potential is obtained by taking the variational derivative of the free energy function,

$$\psi = \frac{\delta\mathcal{F}}{\delta\phi} = \phi^3 - \phi - \varepsilon^2\nabla^2\phi \tag{5}$$

In the phase-field model, the wetting condition is given as

$$\boldsymbol{n} \cdot \nabla\phi = -|\nabla\phi|\cos\theta \tag{6}$$

which is given in (Ding and Spelt, 2007). The default wetting angles at all the domain boundaries in this paper are 90°. Therefore, the corresponding boundary conditions for the phase variable is

$$\partial_n \phi = 0 \tag{7}$$

Besides, assuming the diffusion flux of the phase variable at the boundary to be zero yields the boundary condition of chemical potential as

$$\partial_n \psi = 0 \tag{8}$$

The profile for the flat interface in equilibrium along the normal *z-axis* can be derived as follows

$$\phi(z) = \tanh\left(\frac{z}{\sqrt{2}\varepsilon}\right) \tag{9}$$

The material properties are linearly interpolated by the phase-field function

$$\rho = \frac{\rho_1 + \rho_2}{2} + \frac{\rho_1 - \rho_2}{2}\phi$$

$$\mu = \frac{\mu_1 + \mu_2}{2} + \frac{\mu_1 - \mu_2}{2}\phi \tag{10}$$

The surface tension force in the phase-field method is formulated as (Liu et al., 2014)

$$\vec{f}_s = \frac{3\sqrt{2}}{4}\varepsilon\left[\frac{\sigma}{\varepsilon^2}\psi\nabla\phi + |\nabla\phi|^2\nabla\sigma - (\nabla\sigma \cdot \nabla\phi)\nabla\phi\right] \tag{11}$$

The first term represents the normal surface tension and the last two terms correspond to the Marangoni stress. $\sigma$ represents the constant surface tension coefficient in this article, thus allowing the simplification of the $\vec{f}_s$ as

$$\vec{f}_s = \frac{3\sqrt{2}}{4}\frac{\sigma}{\varepsilon}\psi\nabla\phi \tag{12}$$

## 2.2 Review of classical projection algorithm

In this section, we provide a brief overview of the discrete procedure of the classical projection algorithm (Guermond et al., 2006) for incompressible flow, commonly known as fractional/splitting step methods. In this context, "projection" denotes the process of projecting these velocities onto a space of approximately divergence-free vector fields(Almgren et al., 1998). First split the transient term in Eq. (2) into two parts to produce an intermediate velocity, called $\vec{U}^{n+\frac{1}{2}}$. Then the momentum equation is disassembled into two equations to solve separately. In the actual numerical solution, the process is carried out in the following three steps.

**Classical Step 1: Momentum equation.**

In this step, we consider both convective and diffusive effects, solving Eq. (13) to obtain the intermediate velocity $\vec{U}^{n+\frac{1}{2}}$.

$$\left.\frac{\partial(\rho\vec{U})}{\partial t}\right|_n^{n+\frac{1}{2}} + \nabla\cdot(\rho\vec{U}\otimes\vec{U}) = \nabla\cdot\left[\mu(\nabla\vec{U} + \nabla\vec{U}^T)\right] + \vec{f}_{ext} \tag{13}$$

**Classical Step 2: Pressure/Poisson equation.**

In this step, we account for incompressibility and solve

$$\left.\frac{\partial(\rho\vec{U})}{\partial t}\right|_{n+\frac{1}{2}}^{n+1} = -\nabla P^{n+1} \tag{14}$$

The divergence is computed on both sides of Eq. (14), taking into account the first-order Euler scheme to approximate the time derivative and introduce the continuity condition Eq. (1), we obtain the Poisson equation for the solution of pressure:

$$\nabla \cdot \left(\frac{1}{\rho^{n+1}}\nabla P^{n+1}\right) = \frac{1}{dt}\nabla \cdot \vec{U}^{n+\frac{1}{2}} \tag{15}$$

**Classical Step 3: Final velocity updating.**

Reconsidering the Eq. (14), the intermediate velocity $\vec{U}^{n+\frac{1}{2}}$ and pressure $P^{n+1}$ calculated from the previous two steps are used to update the velocity $\vec{U}^{n+1}$.

$$\vec{U}^{n+1} = -\frac{dt}{\rho^{n+1}}\nabla P^{n+1} + \vec{U}^{n+\frac{1}{2}} \tag{16}$$

## 3 Methodology

In this section, we introduce the non-staggered projection algorithm for two-phase fluid-structure simulation. The Cahn-Hillard equation and the Navier-Stokes equation are solved in an alternating loop until temporal convergence is achieved. The discretization of the phase advection equation Eq.(3) is briefly mentioned, utilizing the second-order TVD Runge-Kutta method in time and WENO5 (Shu and Osher, 1988) for the advection term. Details on this discretization will not be elaborated further. In the subsequent discussion, our primary emphasis lies in the decoupling and modifications of the Navier-Stokes equation Eqs. (1)- (2) under finite volume method (FVM) framework.

### 3.1 Solution of momentum equation

The algorithm presented in this study and the discretization method described in the subsequent section are implemented on two-dimensional non-staggered equidistant Cartesian grids, as illustrated in Figure 1. Nevertheless, the extension to non-equidistant mesh and multiple dimensions is readily achievable.

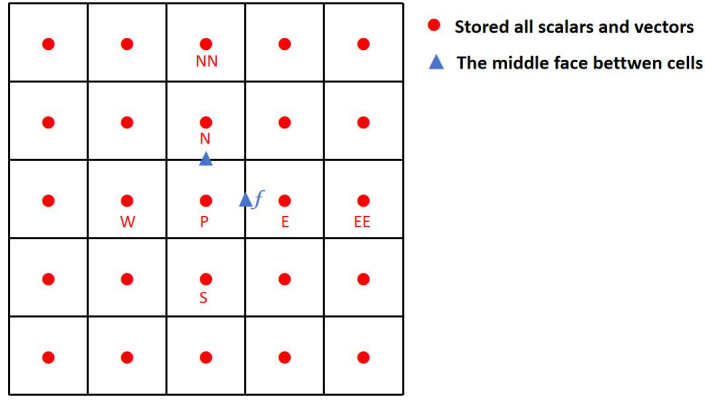

**Figure 1.** The non-staggered equidistant grid, where *f* is the middle face between *P* and *E* cells in the *x*-axis.

The incremental pressure-correction scheme is adopted for decoupling velocity and pressure (Guermond et al., 2006), which is formally similar to the classical projection method, but with the pressure gradient term retained in the momentum equations. The retained term is used to correct the intermediate velocity field and enforce continuity. The momentum equation Eq. (2) is initially decomposed into the following two components,

$$\left.\frac{\partial(\rho \vec{U})}{\partial t}\right|_n^{n+\frac{1}{2}} + \nabla \cdot (\rho \vec{U} \otimes \vec{U}) = \nabla \cdot [\mu(\nabla \vec{U} + \nabla \vec{U}^T)] + \vec{f}_{ext} - \nabla P^n \qquad (17)$$

$$\left.\frac{\partial(\rho \vec{U})}{\partial t}\right|_{n+\frac{1}{2}}^{n+1} = \nabla(-P^{n+1} + P^n) \qquad (18)$$

Use the Backward Difference Formula of second-order (BDF2) to approximate the time derivative in Eq. (17),

$$\left.\frac{\partial(\rho \vec{U})}{\partial t}\right|_n^{n+\frac{1}{2}} = \frac{3\rho^{n+1}\vec{U}^{n+\frac{1}{2}} - 4\rho^n \vec{U}^n + \rho^{n-1}\vec{U}^{n-1}}{2dt} \qquad (19)$$

Simultaneously, the convective term is explicitly treated, while the diffusion term is handled semi-implicitly, resulting in the following expression.

$$\left.\frac{\partial(\rho \vec{U})}{\partial t}\right|_n^{n+\frac{1}{2}} - \nabla \cdot [\mu(\nabla \vec{U})]^{n+\frac{1}{2}} = -\nabla \cdot (\rho \vec{U} \otimes \vec{U})^n + \nabla \cdot [\mu(\nabla \vec{U})^T]^n + \vec{f}_{ext} - \nabla P^n \qquad (20)$$

Discretize Eq. (20) using the finite volume method and solve the corresponding linear matrix with the preconditioned conjugate gradients (PCG) method to obtain the intermediate velocity without Immersed-boundary treatment and Pressure-Oscillation-Free correction, denoted as $\vec{U}^{n+\frac{1}{2}}$.

## 3.2 Immersed-boundary treatment

The boundary conditions for a rectangular computational domain can be readily applied through linear extrapolation. Owing to the employment of a collocated Cartesian grid, the immersion of boundary (IB) technique is easily employed for enforcing the no-slip condition on irregular surfaces. This study specifically embraces the implicit Immersed Boundary Method (IBM) as outlined by Ren (Ren et al., 2012) and Liu et al (Liu and Ding, 2015). Moreover, we employ a simplified region partitioning and optimized calculation of feedback force density to enhance the simplicity and efficiency of the Immersed Boundary Method (IBM) in two-phase flows.

In order to implement no-slip boundary conditions for irregular surfaces, the Immersed Boundary Method (IBM) represents the effect of solid walls on the fluid with a force, denoted as $\vec{f}_{IB}$. However, $\vec{f}_{IB}$ does not directly contribute to the calculation of source terms in the momentum equation; instead, it operates implicitly. Specifically, it generates a corrective velocity $\delta\vec{U}$ that acts on the intermediate velocity $\vec{U}^{n+\frac{1}{2}}$. The corrected velocity $\delta\vec{U}$ is determined through the solution of a system of linear equations, representing a set of Euler-Lagrange systems.

Initially, the fluid-structure boundary is defined by a set of Lagrangian points (totaling *M*), depicted as green squares in Figure 1. The red dots represent background Euler points. The shaded region in the figure corresponds to the adjacent feedback area of the Immersed Boundary (IB) interface, with a total of *N* Euler points falling into this region. In this study, we have set the normal distance between the edge of the feedback area and the IB interface to be 3*h* (where *h* is the Eulerian grid resolution).

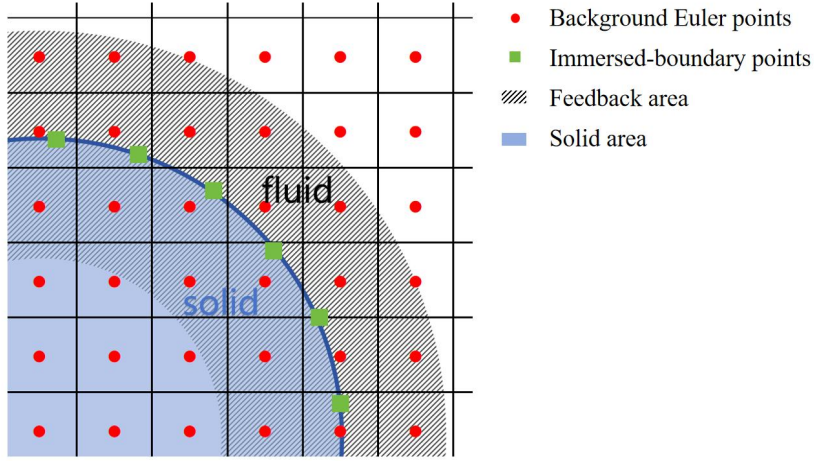

**Figure 2.** Sketch of the Immersed Boundary Method (IBM) used in the present study.

In the second step, employing all Euler points within the vicinity, a Dirac interpolation is applied to each IB point, resulting in a matrix $\boldsymbol{D}$ of dimensions $M \times N$, as

$$\boldsymbol{D} = \begin{bmatrix} D_{11} & D_{12} & \cdots & D_{1N} \\ D_{21} & D_{22} & \cdots & D_{2N} \\ \vdots & \vdots & \ddots & \vdots \\ D_{M1} & D_{M2} & \cdots & D_{MN} \end{bmatrix} \quad (21)$$

$$D_{ij} = \frac{1}{h^2}\delta\left(\frac{x_i^{IB} - x_j^E}{h}\right)\delta\left(\frac{y_i^{IB} - y_j^E}{h}\right), \ (i=1,2,\cdots,M;\ j=1,2,\cdots,N) \quad (22)$$

Where $x$ and $y$ denote spatial coordinates, the superscripts $E$ and $IB$ represent Euler points and IB points, respectively. The specific form of the smooth delta function is given by:

$$\delta(r) = \begin{cases} \dfrac{1}{8}\left(3 - 2|r| + \sqrt{1 + 4|r| - 4r^2}\right) & |r| \leq 1, \\ \dfrac{1}{8}\left(5 - 2|r| - \sqrt{-7 + 12|r| - 4r^2}\right) & 1 < |r| \leq 2, \\ 0 & |r| > 2. \end{cases} \quad (23)$$

Finally, we can derive the following system of linear equations:

$$A = sh^2 \rho_{IB} dt \boldsymbol{D}\boldsymbol{D}^T \quad (24)$$

$$b_x = \begin{bmatrix} \vec{U}_1^{IB} \\ \vec{U}_2^{IB} \\ \vdots \\ \vec{U}_M^{IB} \end{bmatrix} - h^2 \boldsymbol{D} \begin{bmatrix} \vec{U}_1^{n+\frac{1}{2}} \\ \vec{U}_2^{n+\frac{1}{2}} \\ \vdots \\ \vec{U}_N^{n+\frac{1}{2}} \end{bmatrix} \quad (25)$$

$$X = \begin{bmatrix} \vec{a}_1^{IB} \\ \vec{a}_2^{IB} \\ \vdots \\ \vec{a}_M^{IB} \end{bmatrix} \quad (26)$$

where $s=1.2h$ represents the arc length between IB points, and $\rho_{IB}=(\rho_1+\rho_2)/2$. The desired moving velocity of the fluid-structure interface, denoted as $\vec{U}^{IB}$, is set to 0 in the present static study. Solving the linear system above yields the acceleration for each IB point. Subsequently, the IB point accelerations are distributed to the Euler points within the feedback area.

$$\delta\vec{U}_j = s\rho_j dt \sum_{i=1}^{M} \vec{a}_i^{IB} D_{ij}, \quad (j=1,2\cdots,N) \tag{27}$$

And the no-slip correction for the intermediate velocity can be finally expressed as follows:

$$\vec{U}_j^* = \vec{U}_j^{n+\frac{1}{2}} + \delta\vec{U}_j, (j=1,2,\cdots,N) \tag{28}$$

The motivation behind the fluid-structure velocity correction lies in representing the interaction of solid walls with the fluid, achieved through corrective velocity feedback. The use of an implicit algorithm proves more effective in preventing non-physical streamline penetration while ensuring greater tolerance for larger time steps. Furthermore, the coefficient matrix $A$ is solely dependent on the coordinate information of IB points and adjacent Euler points. Therefore, the dimension of $A$ is quite small ($N^2$), and it can be constructed before the iterations of the time step. Consequently, the efficiency and accuracy of the present algorithm are both ensured.

### 3.3 Pressure-Oscillation-Free correction

Subsequently, the Pressure-Oscillation-Free technology in this study is essentially derived from the semi-discrete momentum equation Eq. (20) to further strengthen the pressure–velocity coupling. Substituting velocity with the revised velocity $\vec{U}^*$, which accounts for the consideration and correction of fluid-structure interaction effects. Recall the *x*- component and exclude the source term.

$$\frac{\frac{3}{2}\rho^{n+1}u^*}{dt} - \nabla\cdot(\mu\nabla u^*) = -\nabla_x P^n + \chi \tag{29}$$

where,

$$\chi = \frac{2\rho^n u^n - \frac{1}{2}\rho^{n-1}u^{n-1}}{dt} - \nabla\cdot\left(\rho\vec{U}\otimes u^n\right) + \nabla\cdot[\mu(\nabla u^n)^T] \tag{30}$$

Eq.(29) is further discretized through the finite volume method in the control volume $P$ as shown in Figure 1, and dividing both sides of the discrete momentum equation by the coefficient $a_P$ yields

$$\left(1+\frac{\frac{3}{2}\rho\triangle V}{dt\cdot a_P}\right)u_P^* = \frac{1}{a_P}\sum_{E}^{N}a_i u_i^* - \frac{\triangle V}{a_P}\frac{\partial P}{\partial x}\bigg|_P + \frac{\triangle V}{a_P}\chi_P \qquad (31)$$

The coefficients are defined as:

$$a_P = \frac{\mu_W + 2\mu_P + \mu_E}{2}\frac{\triangle y}{\triangle x} + \frac{\mu_N + 2\mu_P + \mu_S}{2}\frac{\triangle x}{\triangle y}$$

$$a_E = \frac{\mu_E + \mu_P}{2}\frac{\triangle y}{\triangle x}, \quad a_W = \frac{\mu_W + \mu_P}{2}\frac{\triangle y}{\triangle x}$$

$$a_N = \frac{\mu_N + \mu_P}{2}\frac{\triangle x}{\triangle y}, \quad a_S = \frac{\mu_S + \mu_P}{2}\frac{\triangle x}{\triangle y} \qquad (32)$$

where subscript $P$ indicates the cell under consideration, $i$ represents the neighbors of cell $P(E, W, S, N)$, as depicted in Figure 1, and $\triangle V$ is the volume of the cell. The coefficient $a_i$ ($i=P, E, W, S, N$) constitutes the aggregated implicit coefficient encompassing both the transient and viscous terms of the discretized momentum equation. We further reformulate Eq.(31) as

$$(1+cd_P)\,u_P^* = \tilde{u}_P - d_P\frac{\partial P}{\partial x}\bigg|_P + d_P\chi \qquad (33)$$

where

$$d_P = \frac{\triangle V}{a_P}, \quad c = \frac{3\rho}{2dt}$$

$$\tilde{u}_P = \frac{1}{a_P}\sum_E^N a_i u_i^*, \quad i=P,E,W,S,N \qquad (34)$$

The velocity at node $E$ and even the face center $f$ follows the analogous relation as in Eq.(33)

$$(1+cd_E)\,u_E^* = \tilde{u}_E - d_E\frac{\partial P}{\partial x}\bigg|_E + d_E\chi \qquad (35)$$

$$(1+cd_f)\,u_f^* = \tilde{u}_f - d_f\frac{\partial P}{\partial x}\bigg|_f + d_f\chi \qquad (36)$$

The term $u_f^*$ is defined by means of the adjacent cell centers $P$ and $E$ as:

$$\tilde{u}_f = \frac{1}{2}(\tilde{u}_P + \tilde{u}_E) = \frac{1}{2}\left[(1+cd_P)u_P^* + d_P\frac{\partial P}{\partial x}\bigg|_P - d_P\chi_P\right]$$
$$+ \frac{1}{2}\left[(1+cd_E)u_E^* + d_E\frac{\partial P}{\partial x}\bigg|_E - d_E\chi_E\right] \qquad (37)$$

Inserting Eq.(37) in Eq.(36) leads to:

$$(1+cd_f)\,u_f^* = \frac{1+cd_P}{2}u_P^* + \frac{1+cd_E}{2}u_E^* - \left[d_f\frac{\partial P}{\partial x}\bigg|_f - \frac{1}{2}\left(d_P\frac{\partial P}{\partial x}\bigg|_P + d_E\frac{\partial P}{\partial x}\bigg|_E\right)\right]$$
$$+ \left[d_f\chi_f - \frac{1}{2}(d_P\chi_P + d_E\chi_E)\right] \qquad (38)$$

Eq.(38) can be further simplified by approximation

$$1 + cd_f = 1 + cd_P = 1 + cd_E$$
$$d_f = d_P = d_E \tag{39}$$

and

$$\hat{d}_f = \frac{d_f}{1 + cd_f} \tag{40}$$

Inserting Eqs.(39) and (40) to Eq.(38) yields

$$u_f^* = \frac{u_P^* + u_E^*}{2} - \hat{d}_f \left[ \left.\frac{\partial P}{\partial x}\right|_f - \frac{1}{2}\left( \left.\frac{\partial P}{\partial x}\right|_P + \left.\frac{\partial P}{\partial x}\right|_E \right) \right] + \hat{d}_f \left[ \chi_f - \frac{\chi_P + \chi_E}{2} \right] \tag{41}$$

Replace the differential of pressure with the central difference, and consider $\chi_f = (\chi_P + \chi_E)/2$. Eq.(41) becomes

$$u_f^* = \frac{u_P^* + u_E^*}{2} - \frac{\hat{d}_f}{4\Delta x}[P_W - 3P_P + 3P_E - P_{EE}] \tag{42}$$

Convection velocity $u_f^*$ calculated from a compact pressure stencil is the key to damping out the unphysical pressure oscillations (Denner and van Wachem, 2014; Kawaguchi et al., 2002), because

$$\frac{\hat{d}_f}{4\Delta x}[P_W - 3P_P + 3P_E - P_{EE}] \sim \frac{\partial^3 P}{\partial x^3} \tag{43}$$

The velocity $u_f^*$ above is suitable for the single-phase flow. For multiphase flows, the variable density and the surface force have to be considered for the velocity correction, because neglecting surface forces can lead to substantial imbalances (Denner and van Wachem, 2014).

The harmonic average provides a meaningful interpolation of the face density and reduces the impact of large density ratios (Ferziger, 2003)

$$\rho_f = \frac{2}{\rho_P^{-1} + \rho_E^{-1}} \tag{44}$$

Previous work (Guo, 2021; Poblador-Ibanez and Sirignano, 2022) has established that the surface force causes a pressure jump in the two-phase flow, and therefore, the surface force should be included in the momentum interpolation method (Denner and van Wachem, 2014). By combining with the surface tension model in Eq.(12), and applying the same stencil for pressure, Eq.(42) is modified for multiphase flow, and the corrected intermediate velocity is given by

$$\hat{u}_f = \frac{u_P^* + u_E^*}{2} - \frac{\hat{d}_f}{4\Delta x}[P_W - 3P_P + 3P_E - P_{EE}] \\ + \frac{3\sqrt{2}}{4}\frac{\sigma}{\varepsilon}\left[ \psi_f \left.\frac{\partial \phi}{\partial x}\right|_f - \frac{\rho_f}{2}\left( \frac{\psi_P}{\rho_P}\left.\frac{\partial \phi}{\partial x}\right|_P + \frac{\psi_E}{\rho_E}\left.\frac{\partial \phi}{\partial x}\right|_E \right) \right] \tag{45}$$

The motivation behind the density weighting is to dampen pressure oscillations potentially arising from large density jumps at the interface. Eq.(45) provides an equation to compute the advective velocity at face centers for two-phase flows using the phase-field method and can be easily extended to other axes. Utilizing spline interpolation, the corrected face-centered velocities are interpolated back to the cell centers for subsequent calculations.

### 3.4 Solution of Poisson equation

The subsequent work is updating pressure $P^{n+1}$. Take the divergence on both sides of Eq. (18) and consider the continuity condition Eq.(1). Thus, the Poisson equation is obtained:

$$\nabla \cdot \left( \frac{1}{\rho^{n+1}} \nabla P' \right) = \frac{3}{2dt} \nabla \cdot \widehat{U} \tag{46}$$

where $\widehat{U}$ is the modified intermediate velocity based Eq. (45) and $P^{n+1} = P' + P^n$. The linear system resulting from Eq. (46) is solved using a fast Fourier transform (FFT) based Poisson (Helmholtz) solver.

### 3.5 Final velocity updating

Finally, the incompressible velocity field $\vec{U}^{n+1}$ can be obtained from Eq. (18), yielding,

$$\vec{U}^{n+1} = \vec{U}^* - \frac{2dt}{3\rho^{n+1}} \nabla P' \tag{47}$$

### 3.6 Algorithm overview

Thus far, a complete time step for simulating the incompressible two-phase fluid-structure interactions on non-staggered grid has been introduced, requiring only two simple modifications inside the standard procedure. The algorithm for advancing from time step *n* to *n+1* is summarized in the following six-step procedure:

**Step 1.** Advance the evolution of phase interfaces by employing the second-order Runge-Kutta (RK2) algorithm to solve the Cahn-Hilliard equation. Specifically, update the phase fraction $\phi^{n+1}$ and reconstruct the density and viscosity distributions for time step *n* to *n+1*.

**Step 2.** Calculate the uncorrected velocity $\vec{U}^{n+\frac{1}{2}}$ by solving the momentum equation Eq. (20), utilizing the BDF2 scheme for transient advancement and fifth-order WENO for the convective term.

**Step 3.** Calculate the acceleration of Immersed Boundary (IB) points through the linear system by Eq. (24)-(28). And add the no-slip correction on the intermediate velocity $\vec{U}^{n+\frac{1}{2}}$ by Eq. (27)-(28), resulting in $\vec{U}^*$.

**Step 4.** Motivate the transient velocity $\vec{U}^*$ for suppressing the unphysical pressure oscillation through the application of Eq. (45), resulting in $\widehat{U}$.

**Step 5.** Update the pressure field $P^{n+1}$ by utilizing the corrected transient velocity $\widehat{U}$ obtained in Step 4 through Eq. (46).

**Step 6.** Advance the velocity field to *n+1* by Eq. (47), based on the corrected transient velocity $\vec{U}^*$ from Step 3 and the obtained pressure $P'$ from Step 5.

Note that the assembled linear system of Step 2 and Step 5 is solved through the Preconditioned Conjugate Gradient (PCG) algorithm and the fast Fourier transform (FFT) based Poisson (Helmholtz) solver, respectively. Step 3 is optional and depends on whether fluid-structure interaction effects are being considered.

## 4 Results & Discussion

### 4.1 Reversed single vortex

The reversed single vortex, evolving in a given background shear flow, is a benchmarking example for two-phase flow computation introduced by Bell (Bell et al., 1989) and first applied by Rider and Kothe (Rider and Kothe, 1998). Following the case set up in (Cho et al., 2011), we initially place a circular fluid disk with a radius of 0.15 at (0.5, 0.75) in a square computational domain of [0, 1] × [0, 1]. The velocity is defined by the stream function:

$$\Psi = \frac{1}{\pi}\sin^2(\pi x)\sin^2(\pi y)\cos\left(\frac{\pi t}{T}\right) \tag{48}$$

The velocity components can then be derived as:

$$\begin{aligned} u &= -\frac{\partial \Psi}{\partial x} = -2\sin^2(\pi x)\sin(\pi y)\cos(\pi y)\cos\left(\frac{\pi t}{T}\right) \\ v &= -\frac{\partial \Psi}{\partial y} = 2\sin^2(\pi y)\sin(\pi x)\cos(\pi x)\cos\left(\frac{\pi t}{T}\right) \end{aligned} \tag{49}$$

where $t$ and $T$ are the evolution time and period, respectively. With the contribution of the $\cos(\pi t/T)$ term, the circle is fully stretched at $t = T/2$ and returns to its initial state at $t = T$.

In this study, computations with a period of $T = 2$ based on a mesh of 32×32, 64×64, 128×128 and 256×256 grid points are carried out. The CFL number and mobility are fixed at 0.1 and 0.0001 respectively. To investigate the spatial order of convergence, the relative position error is defined as:

$$Er = \frac{\sum_{i=1}^{N} |\phi_i^{final} - \phi_i^0|}{\sum_{i=1}^{N} |\phi_i^0|} \tag{50}$$

$N$ represents the total number of all nodes, and the superscripts of $\phi$ denote the initial moment and one period, respectively.

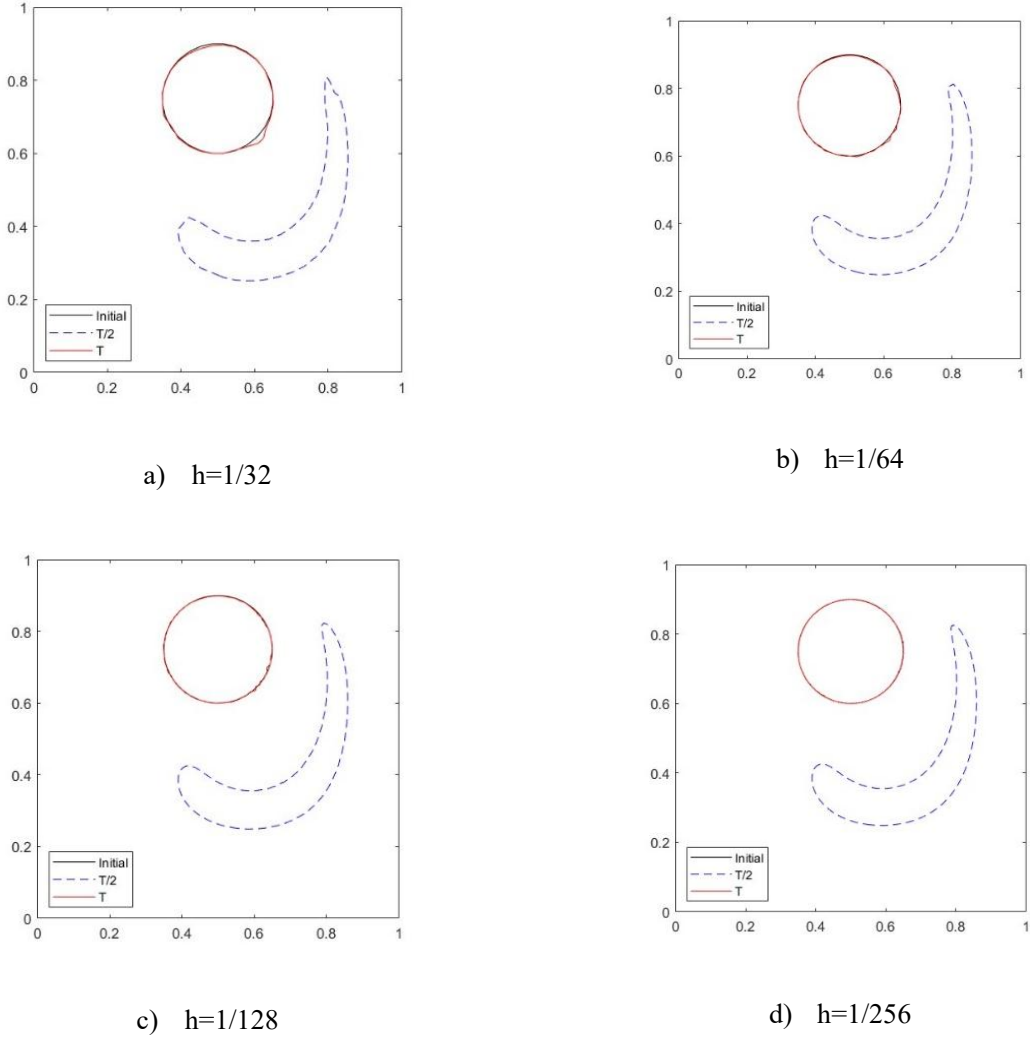

**Figure 3.** Results of the reversed single vortex problem. Black thick line: at $t = 0$; blue dashed line: at $t=T/2$; red dash-dotted line: at $t=T$

It is clearly demonstrated that the results from the present algorithm converge to the exact solution with a finer mesh, as shown in Figure 3. Table I presents the quantitative relative position errors, defined as the discrepancy between the initial and final positions of the interface profiles following Eq.(50).

**Table 1.** The errors of the reversed single vortex problem are shown in the table.

| Mesh size | 1/32 | 1/64 | 1/128 | 1/256 |
|---|---|---|---|---|
| Errors | 3.25e-02 | 1.54e-2 | 9.04e-3 | 4.50e-3 |

## 4.2 Lid driving cavity

The lid-driven cavity flow is first conducted, which has been established as a standard benchmark test for numerical methods of incompressible fluid dynamics (Ghia et al., 1982; Zhang et al., 2014). In two-phase flow, the density and viscosity changes following the interfacial dynamics as Eq.(10), but remain constant in single-phase flow. Therefore, the proposed scheme for a two-phase flow system should recover the results of a single-phase flow system when there is only one fluid. To verify the reduction consistency, we perform three cases of Reynolds numbers 100, 1000, and 3200 with the 64×64, 100×100, and 200×200 grid respectively. The profiles of *x*-component *u* along the vertical centerline and *y*- component v along the horizontal centerline are plotted in Figure 4, and results from the present computations accord well with the reference solutions in (Ghia et al., 1982).

Moreover, to demonstrate the strength of the present method in avoiding pressure oscillation, the present algorithm is compared with the classical projection method based on the Re=1000 case (Guermond et al., 2006). A smooth pressure field can be predicted and no odd-even decoupling occurs in the present computations, while clear pressure oscillations exist in the results from the classical projection method in Figure 5.

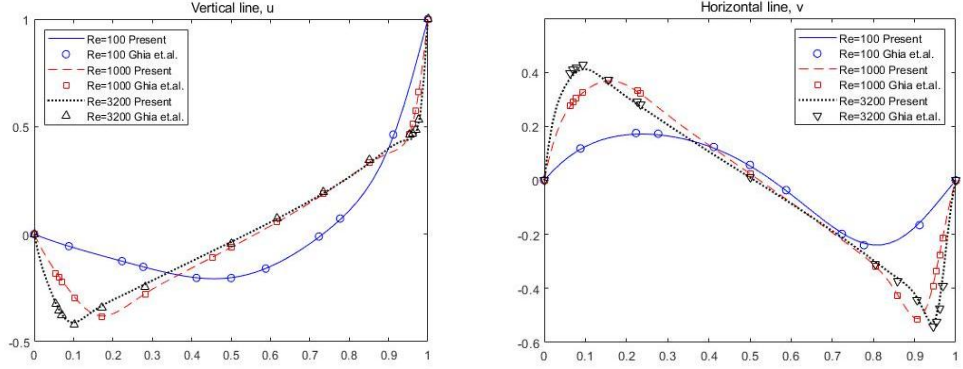

**Figure 4.** Numerical results of the driven cavity. Left u-velocity at the vertical centerline. Right: v-velocity at the horizontal centerline. Ghia et al.'s results Circle: Re = 100, Square: Re = 1000, Diamond: Re = 3200. Present Solid line: Re = 100, Dash line: R

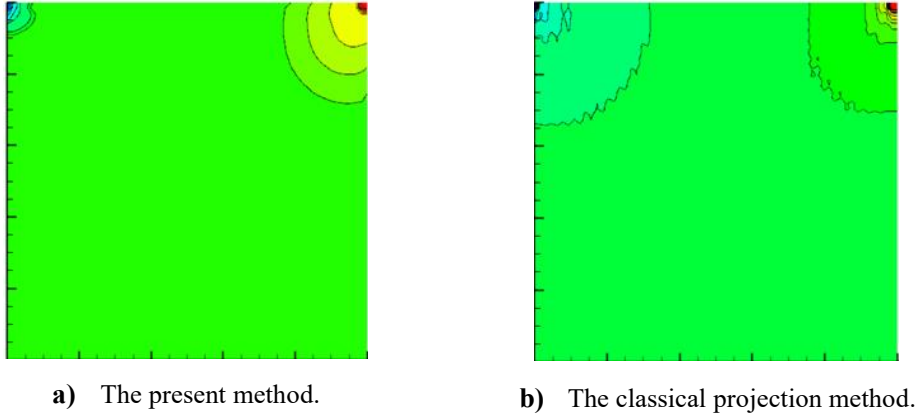

| a) The present method. | b) The classical projection method. |

**Figure 5.** The pressure contours for Re=1000 on 128×128 grids. Computational results are obtained using the present method and the classical projection method.

### 4.3 Droplet dynamic

The droplet dynamic behavior is a popular and challenging task in the numerical study of multiphase flow. In this section, we first simulated a standard stationary droplet without gravity (Abadie et al., 2015; Huang et al., 2019) to test the surface tension and its convergence. Secondly, to further test the effect of the present algorithm on suppressing pressure oscillation in two-phase flow, we introduce the gravity effects to induce droplet rise due to buoyancy.

### 4.3.1 Stationary/ buoyancy droplet

In this beginning test, we set a droplet with a diameter $D = 0.4m$ in a 1m×1m domain without gravity. The free-slip boundary condition is applied to all the boundaries, and the initial velocity is zero everywhere. The stationary drop and the surrounding fluid have the same densities and viscosities. The dimensionless parameter that characterizes this problem is the Laplace number $La = \sigma\rho D/\mu^2$. To vary $La$, the density is varied for different tests while other parameters including surface tension coefficient ($\sigma = 1\text{N/m}$) and viscosity ($\mu = 0.1\text{Pa}\cdot\text{s}$) remain constant. All simulations are stopped at t=10s.

According to Laplace's law, the stationary drop remains in equilibrium under the action of surface tension, and the pressure difference between internal and external is theoretically determined by $\Delta P = n\sigma/R$. $n$ equals 1 for two-dimensional problems, and for two-dimensional axisymmetric and three-dimensional problems, $n=2$. As shown in Figure 6, we test three different $La$ numbers (120, 1200, 12000) on a 128×128 grid. The relative errors between numerical solutions and theoretical predictions are about 1%, which confirms the accuracy of the present method.

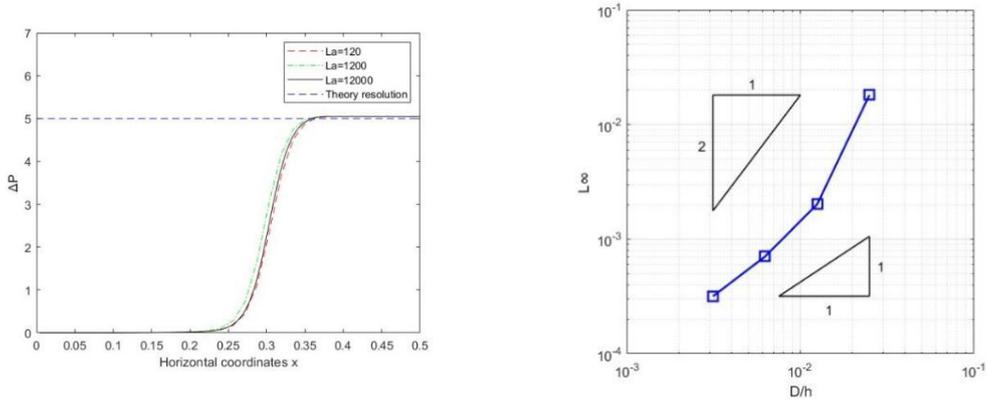

**Figure 6.** Left: The pressure distribution on the horizontal midline, the corresponding errors of La=120, 1200, and 12000 are 1.011 %, 1.035%, and 1.04% respectively. Right: Convergence study on the accuracy of pressure/surface tension force calculation.

In practice, depending on the method used for discretizing the surface-tension force and the pressure gradient, an exact numerical balance is difficult to obtain and often leads to so-called "spurious" or ''parasitic'' velocities fed by this imbalance (Inguva et al., 2022; Kang et al., n.d.). We use the magnitude of the resultant velocity, as the local strength of the spurious current. Figure 6 illustrates the evolution of the spurious current with grid refinement (the cell size h is 1/16, 1/32, 1/64, 1/128, respectively), and the convergent rate is between the 1st and 2nd order. The result is in accordance with the predictions of Magnini et.al(Magnini et al., 2016), which showed that the magnitude of the maximum spurious current in the domain should have the same convergence rate as the curvature evaluation as the mesh is refined.

In two-phase flow, pressure oscillation may lead to poor results. The 'wiggles' at the pressure reference point could generate a progressive pressure gradient, especially in long-time simulation, which brings unphysical behavior of the fluid.

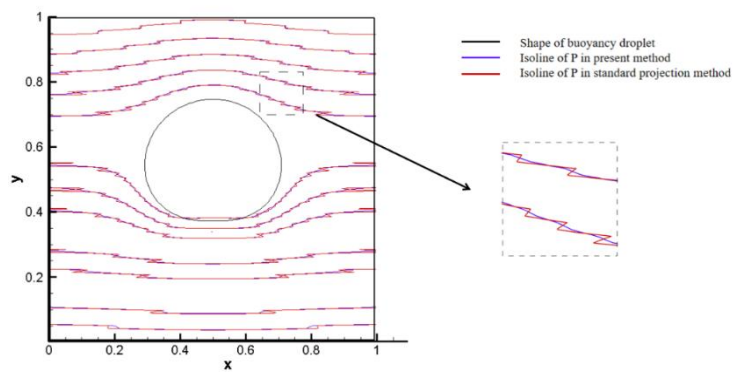

**Figure 7.** Comparison study of pressure oscillation between the classical projection method and the present method.

To verify the performance of the present model in two-phase flow, we added gravity in the above stationary case, which causes the buoyancy rise of the stationary droplet. The density ratio of the droplet to the surrounding liquid is 1:1000, the viscosity ratio is 0.1:1, and the gravity is set at 9.8m/s$^2$. The calculation was still carried out on 128×128 grids. Figure 7 shows the shape of the droplet and the pressure contour at t=0.125s. From the enlarged portions of the contour, it can be observed that the pressure curve from the present method is smoother, while the classical projection method shows a jagged shape clearly. This result demonstrates that the algorithm has excellent Pressure-Oscillation-Free performance even in two-phase flow with a large density ratio.

**4.3.2 Droplet in the shear flow**

Understanding drop deformation is crucial for designing industrial and environmental fluid devices. While previous research has mainly focused on equal-density ratios in two-dimensional viscous drop deformation in shear flow, investigating drops with high-density ratios is both essential and challenging. Simulating the deformation of high-density ratio droplets in shear flow on a non-staggered grid, non-physical pressure oscillations can arise from incorrect odd-even decoupling of velocity and pressure, leading to the evolution of a source or sink that drives unexpected movement of the two-phase interface. The oscillation-free algorithm employed in this study effectively suppresses non-physical oscillations, resulting in more reasonable droplet deformation outcomes.

In shear flow, a droplet undergoes stretching and tank-treading motion. The setup involves confining the flow domain between two parallel walls as shown in Figure 8, with a circular droplet initially placed at the center. The domain is [2×2], and the droplet radius is 0.5. The upper and lower boundaries are no-slip walls and move at velocities of 0.1 and -0.1, respectively, while the left and right boundaries are periodic. The grid comprises $128^2$ nodes, and a time step of 1e-4 is utilized throughout the simulation. Set the physical property parameters according to Table 2, for validating the performance of the present algorithm in large density ratio two-phase flow.

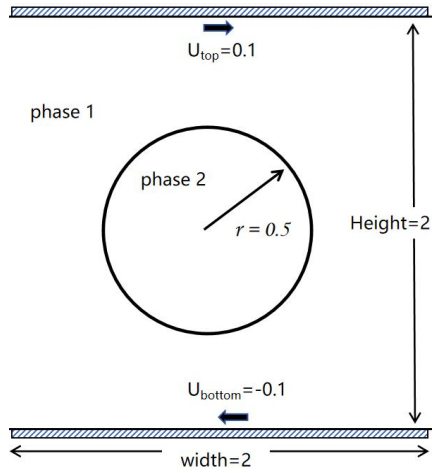

**Figure 8.** Sketch of the initial set of droplet deformation in shear flow

**Table 2.** Physical property parameters of large-ratio-density droplet deformation simulation in shear flow

| Property | $\mu_1$ | $\mu_2$ | $\sigma$ | $\rho_1$ | $\rho_2$ |
|---|---|---|---|---|---|
| Case 1 | 0.1 | 0.1 | 0.01 | 1 | 10 |
| Case 2 | 0.1 | 0.1 | 0.01 | 1 | 100 |

Figure 9 illustrates the droplet deformation in shear flow at a density ratio of 10. In the absence of Pressure-Oscillation-Free correction (red solid line), the droplet is driven away by unphysical pressure oscillations. However, the proposed algorithm (dashed black line) ensures the stability of the droplet centroid over long-term simulations, consistent with previous numerical and experimental studies. Figure 10 depicts the simulation with a density ratio of 100. In the absence of correction, numerical errors arise at 2.3 seconds due to pressure oscillations, resulting in program termination, whereas the present algorithm demonstrates robustness in simulations extending up to 100s.

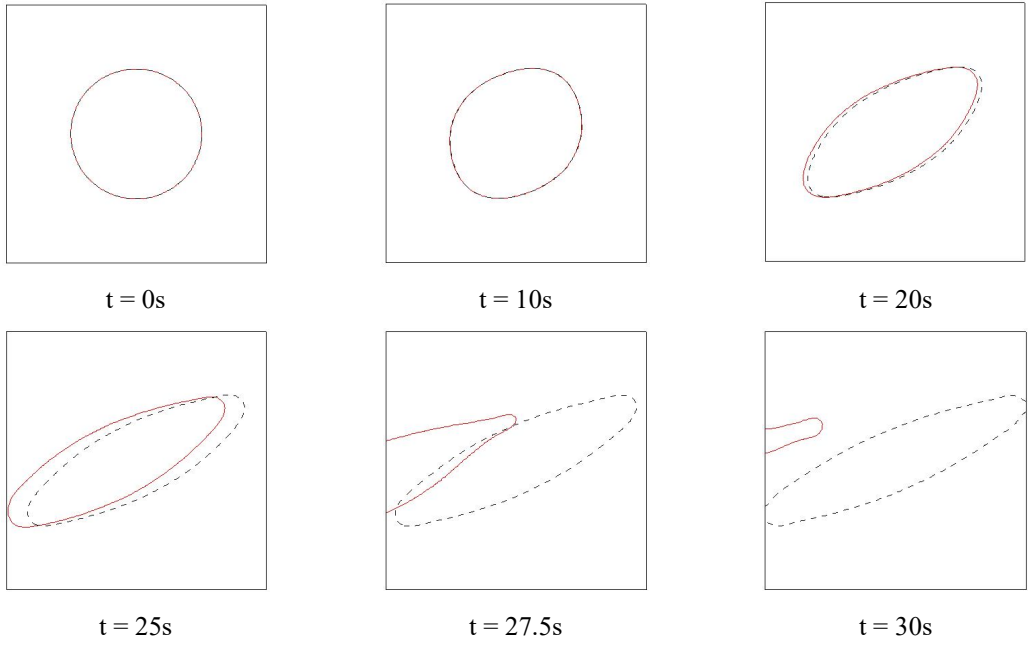

**Figure 9.** Diagram of large density ratio droplet deformation in shear flow, comparison of Pressure-Oscillation-Free correction (black dashed line) and classical projection algorithm (red solid line)

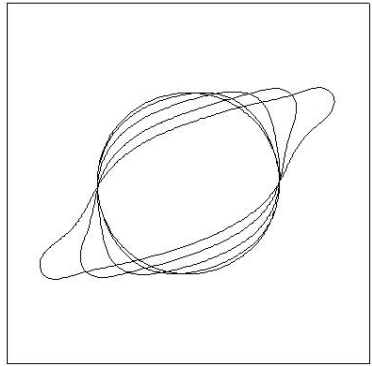

**Figure 10.** Metamorphosis history of large density ratio (=100) droplet deformation in shear flow by present non-staggered projection algorithm. Black lines extending from the center outwards represent t=20s, 40s, 60s, 80s, 100s.

## 5 Numerical results

### 5.1 Rayleigh-Taylor instability

In this section, we model the Rayleigh-Taylor instability, a benchmark of immiscible two-phase flow involving complex topological interface evolution. The computational domain is [1×4] and a uniform grid of 128×512 is used. The no-slip boundary conditions are applied at the top and the bottom boundaries while the symmetric boundary condition is imposed at the two vertical boundaries. The denser fluid is initially placed above the lighter with a horizontal interface, and the velocity is everywhere zero. The phase variable is initially distributed as follows:

$$\phi(t=0) = \tanh\left(\frac{1}{\sqrt{2}\,\varepsilon}(y + 0.1\cos(2\pi x))\right) \tag{51}$$

Due to gravity, the heavy fluid sinks, penetrating through the fluid below. The dimensionless number of this example is the Atwood number $At = (\rho_H - \rho_L)/(\rho_H + \rho_L)$ and the Reynolds number $Re = \rho U L/\mu$. We simulated the case of At=0.5 and Re=256. Figure 11 shows the shape of the interface. The denser liquid starts out going down, and the lighter liquid goes up. Small-scale perturbations appear and begin to grow on the side of the descending column of the denser fluid, resulting in a mushroom-like interface. Over time, the tail became very thin and the topology changed, resulting in very small structural patterns. Figure 12 shows the quantitative comparison with other literature. It can be seen that the results obtained here are in good agreement with the reference results in He and Xiao et.al (He et al., 1999; Xiao et al., 2022a).

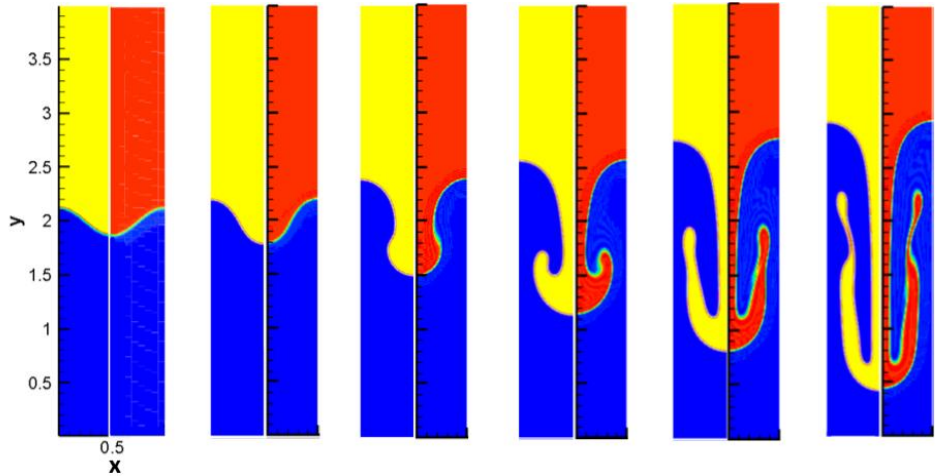

**Figure 11.** Flow patterns of the Rayleigh-Taylor instability with At=0.5, Re=256, at t=0.5, 1, 2, 3, 4, 5s, The left yellow interface corresponds to the presented algorithm, and the right red interface represents Xiao et al (Xiao et al., 2022a).

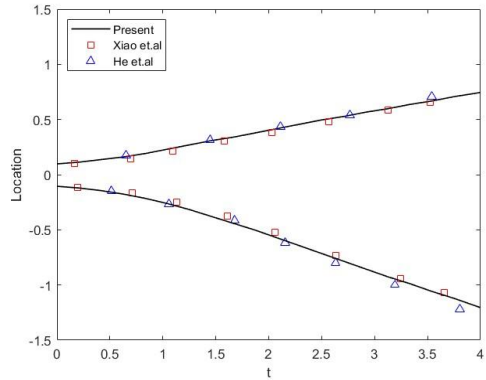

**Figure 12.** The transient locations of the tips of bubbles going up and down, At=0.5, Re=256.

## 5.2 Coalescence of bubbles

We simulated the process of droplet merge to verify the performance of this algorithm in suppressing mass leakage. At the initial moment, two circular droplets are placed in a square area of 1m×1m, the resolution of the uniform grids is 200×200, and symmetric boundary conditions are applied all around. The diameter of the two droplets is 0.3m, the initial separation distance D is 0.02m, and the thickness of the interface is 0.04m. Since the initial distance of droplets is smaller than the thickness of the interface, the static droplets will merge under the action of surface tension (Zheng et al., 2006). In addition, the density ratio of the surrounding fluid to the bubble is 1:1000.

The interface evolution process is shown in Figure 13. The present method can accurately capture the process from contact to the merge of two droplets until the formation of a new circular droplet, which is coincident with previous relevant works (Zheng et al., 2006). To quantitatively evaluate the mass conservation properties of the proposed method, the resultant droplet size at a steady state is compared with the analytical solution. The error of the present prediction is 1.14%. While the error of the original phase-field method reaches up to 8.42% from the comparison experiments by Zhang et.al (Zhang et al., 2019).

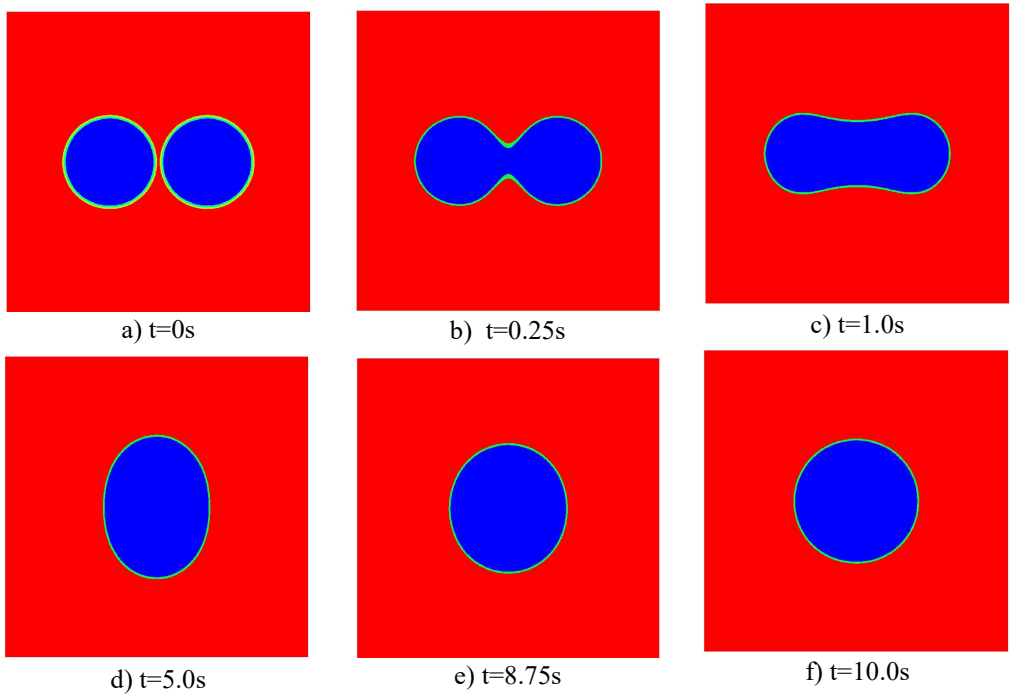

**Figure 13.** Evolution of the interface shape of two merging bubbles. From left to right and from top to bottom, t = 0, 0.25, 1.0, 5.0, 8.75, 10.0 (s).

## 5.3 Rising bubble in rectangular/complex domain

Hysing et al (Hysing et al., 2009) published a pure numerical benchmark with two test cases for a 2D rising bubble in a rectangular domain, which is widely adopted for testing in two-phase flow algorithm(Klostermann et al., 2013; Sharma et al., 2021; Xiao et al., 2022b) and considered here. The bubble is initially located at (x, y) = (0.5,0.5) with D= 0.5 as the initial diameter. The viscosity and density of fluid 1 (bubble) are smaller than those of the surrounding fluid 2. The domain is a fully enclosed cavity with length [1×2] by no-slip walls at the top and the bottom and free-slip walls on the left and the right. The gravity vector points toward the bottom of the domain. The physical parameters are given in Table 3. This case uses a fixed setup with a grid of 64×128, a time step of 1e-4, and an interface width of 0.02. The bubbles rise slowly due to buoyancy, and Figure 14 shows the evolution of the bubble shape. The shape based on the present algorithm matches the reference solution (Xiao et al., 2022b).

We use two benchmark quantities: the center of mass $y_c$ and the rising velocity $v_c$ to quantify our results same as in (Xiao et al., 2022b).

$$y_c = \frac{\int_\Omega y \frac{1+\phi}{2} d\Omega}{\int_\Omega \frac{1+\phi}{2} d\Omega} \tag{52}$$

$$v_c = \frac{\int_\Omega v \frac{1+\phi}{2} d\Omega}{\int_\Omega \frac{1+\phi}{2} d\Omega} \tag{53}$$

where $y$ is the vertical coordinate value and $v$ is the y-component of the velocity $U$. The results from the present method match well with the reported data (Xiao et al., 2022b) as in Figure 15.

**Table 3.** The physical parameters in the bubble rising test.

| Parameters | $\rho_1$ | $\rho_2$ | $\mu_1$ | $\mu_2$ | g | σ |
|---|---|---|---|---|---|---|
| Value | 100 | 1000 | 1 | 10 | 0.98 | 24.5 |

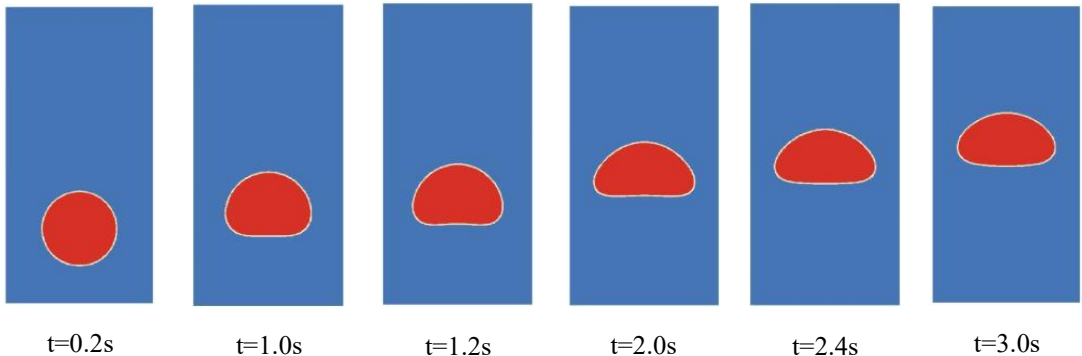

t=0.2s   t=1.0s   t=1.2s   t=2.0s   t=2.4s   t=3.0s

**Figure 14.** Shape evolution of rising bubble from t=0s to t=3s, simulated on a grid of 64×128.

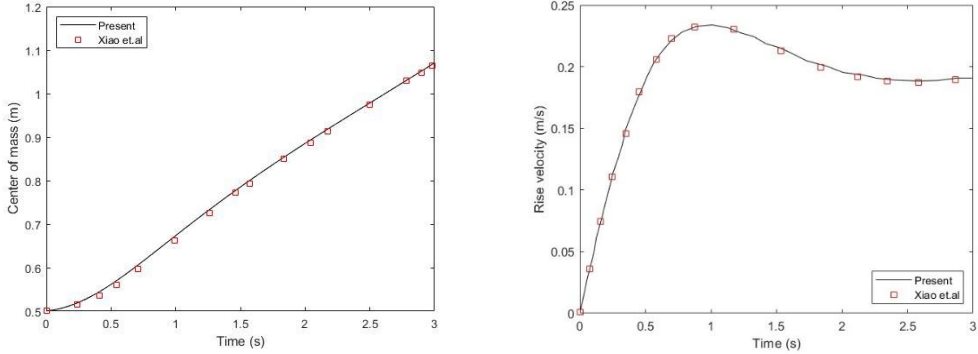

**Figure 15.** Comparison of the mass center position $y_c$ and the rise velocity between the present algorithm and high-order spectral element method (Xiao et al., 2022b).

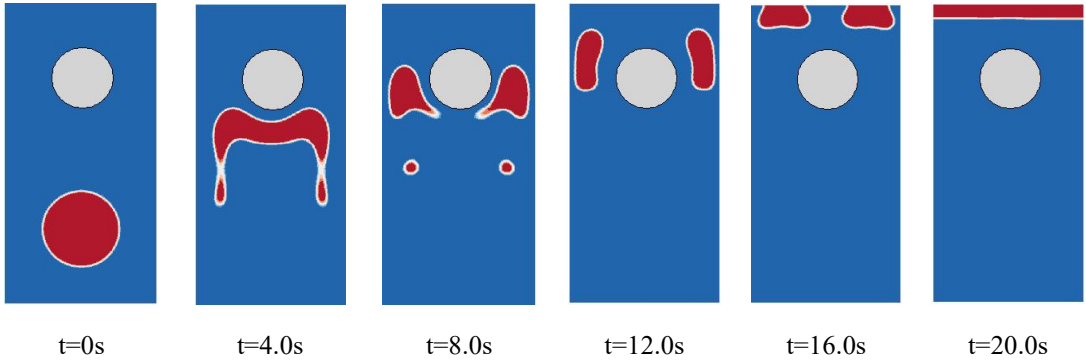

| t=0s | t=4.0s | t=8.0s | t=12.0s | t=16.0s | t=20.0s |

**Figure 16.** Shape evolution of rising bubble simulation from t=0s to t=20s with a cylindrical obstacle in the channel simulated on a grid of 64×128.

Next, we further validate the proposed two-phase fluid-structure coupling algorithm by considering a bubble-rising case with a cylindrical obstacle in the channel. A stationary cylinder with a radius of 0.2 is positioned at the spatial coordinates [0.5, 1.5]. As the bubble ascends, it undergoes splitting due to the fluid-structure coupling effect, eventually merging at the top. The simulation results in Figure 16 demonstrate that the algorithm presented in this study effectively simulates two-phase flow with complex wall surfaces. The qualitative aspects of the interface are captured well. This finding holds significance for engineering applications, particularly in scenarios involving obstacles within fluid channels or pipelines. Such situations are common in various engineering and scientific domains, including oil and gas transportation, chemical processing, and environmental engineering. Therefore, the algorithm's ability to accurately simulate the flow behavior around obstacles contributes practically to optimizing pipeline design and fluid control.

It is noteworthy that the results in this section demonstrate the effectiveness of the algorithm in handling flow problems with complex geometries. However, despite the simulation results aligning with those of the high-order spectral element method reported in the literature, further research and validation are still warranted. Future research efforts could focus on exploring the influence of different types of obstacles on fluid dynamics behavior and optimizing the algorithm for improved computational efficiency and accuracy. These endeavors will contribute to expanding the algorithm's applicability in engineering practice and advancing the field of fluid dynamics simulation technology.

## 6 Conclusion

In this paper, we introduce a modified projection algorithm for simulating large-density-ratio multiphase fluid-structure interactions on a non-staggered Cartesian grid. The conventional projection algorithm, typically employed for incompressible Navier-Stokes equations, sequentially solves for intermediate velocities, pressures, and final velocities. Our modified projection algorithm introduces two additional steps within this process. The first modified step enforces the non-slip condition at fluid-structure interfaces. The second one addresses pressure oscillations arising from non-staggered grids. The fluid-fluid interface is characterized by the Cahn-Hilliard equation. Temporal discretization employs a second-order Runge-Kutta method, while spatial discretization utilizes a fifth-order Weighted Essentially Non-Oscillatory scheme for convection terms. Particularly, we construct the velocity at cell faces using a momentum-weighted interpolation method integrated with the phase-field surface tension model. The combination of the two is a novelty of the present work. This ensures a compact pressure-velocity coupling and maintains a discrete balance between pressure gradient and surface tension. Implementation of the non-slip condition for curved surfaces within the flow field is achieved through the implicit Immersed Boundary Method (IBM). For static fluid-structure boundaries, solving for feedback acceleration by a relatively small linear equation system with a steady coefficient matrix is sufficient.

The verification studies section presented three cases to validate the accuracy and effectiveness. Firstly, the reversed single vortex and lid-driven cavity flow cases were employed to verify the accuracy of the two-phase interface capturing and the decoupling for incompressible flow fields, respectively. Subsequently, in the droplet dynamics section, three distinct scenarios were examined. Initially, a stationary droplet without gravity was simulated to evaluate surface tension convergence. Subsequently, gravity effects were introduced to observe droplet behavior due to buoyancy, showcasing the algorithm's capability to suppress pressure oscillations even under conditions of large-density-ratio flow. Lastly, droplet deformation in shear flow was investigated, demonstrating the algorithm's ability to effectively suppress unphysical oscillations and produce realistic droplet deformation outcomes. The numerical results section validates our algorithm through three key simulations: Rayleigh-Taylor instability, bubble coalescence, and rising bubbles with and without obstacles. These simulations provided compelling evidence of the algorithm's stability and reliability in addressing various challenging multiphase fluid-structure coupling scenarios, highlighting its innovative contributions to the field of fluid dynamics research and engineering.

Overall, owing to the utilization of non-staggered grids and the low-cost corrective steps, the algorithm proposed in this paper demonstrates remarkable scalability. However, it is important to acknowledge the limitations of the current algorithm and potential areas for improvement. Future research endeavors will explore extensions to three-dimensional and moving boundary problems, with a continued focus on enhancing computational efficiency.

## Declaration of competing interest

The authors declare that they have no known competing financial interests or personal relationships that could have appeared to influence the work reported in this paper.

## Acknowledgments

This work is supported by the National Natural Science Foundation of China (No. 12172070, 12102071), and the Fundamental Research Funds for the Central Universities (No.2021CDJQY-055).

## References

Abadie, T., Aubin, J., Legendre, D., 2015. On the combined effects of surface tension force calculation and interface advection on spurious currents within Volume of Fluid and Level Set frameworks. Journal of Computational Physics 297, 611–636. https://doi.org/10.1016/j.jcp.2015.04.054

Almgren, A.S., Bell, J.B., Colella, P., Howell, L.H., Welcome, M.L., 1998. A Conservative Adaptive Projection Method for the Variable Density Incompressible Navier–Stokes

Equations. Journal of Computational Physics 142, 1–46. https://doi.org/10.1006/jcph.1998.5890

Armfield, S.W., Street, R., 2000. Fractional step methods for the Navier-Stokes equations on non-staggered grids. ANZIAMJ 42, 134. https://doi.org/10.21914/anziamj.v42i0.593

Bell, J.B., Colella, P., Glaz, H.M., 1989. A second-order projection method for the incompressible navier-stokes equations. Journal of Computational Physics 85, 257–283. https://doi.org/10.1016/0021-9991(89)90151-4

Bourlioux, A., 1995. A coupled level-set volume-of-fluid algorithm for tracking material interfaces, in: Proceedings of the 6th International Symposium on Computational Fluid Dynamics, Lake Tahoe, CA.

Cho, M.H., Choi, H.G., Yoo, J.Y., 2011. A direct reinitialization approach of level-set/splitting finite element method for simulating incompressible two-phase flows. Numerical Methods in Fluids 67, 1637–1654. https://doi.org/10.1002/fld.2437

Denner, F., van Wachem, B.G.M., 2014. Fully-Coupled Balanced-Force VOF Framework for Arbitrary Meshes with Least-Squares Curvature Evaluation from Volume Fractions. Numerical Heat Transfer, Part B: Fundamentals 65, 218–255. https://doi.org/10.1080/10407790.2013.849996

Ding, H., Spelt, P.D.M., 2007. Wetting condition in diffuse interface simulations of contact line motion. Phys. Rev. E 75, 046708. https://doi.org/10.1103/PhysRevE.75.046708

Ding, H., Spelt, P.D.M., Shu, C., 2007. Diffuse interface model for incompressible two-phase flows with large density ratios. Journal of Computational Physics 226, 2078–2095. https://doi.org/10.1016/j.jcp.2007.06.028

Duy, T.-N., Nguyen, V.-T., Phan, T.-H., Park, W.-G., 2021. An enhancement of coupling method for interface computations in incompressible two-phase flows. Computers & Fluids 214, 104763. https://doi.org/10.1016/j.compfluid.2020.104763

Elahi, R., Passandideh-Fard, M., Javanshir, A., 2015. Simulation of liquid sloshing in 2D containers using the volume of fluid method. Ocean Engineering 96, 226–244. https://doi.org/10.1016/j.oceaneng.2014.12.022

Ferziger, J.H., 2003. Interfacial transfer in Tryggvason's method. Int. J. Numer. Meth. Fluids 41, 551–560. https://doi.org/10.1002/fld.455

Francois, M.M., Cummins, S.J., Dendy, E.D., Kothe, D.B., Sicilian, J.M., Williams, M.W., 2006. A balanced-force algorithm for continuous and sharp interfacial surface tension models within a volume tracking framework. Journal of Computational Physics 213, 141–173. https://doi.org/10.1016/j.jcp.2005.08.004

Galusinski, C., Vigneaux, P., 2008. On stability condition for bifluid flows with surface tension: Application to microfluidics. Journal of Computational Physics 227, 6140–6164. https://doi.org/10.1016/j.jcp.2008.02.023

Ghia, U., Ghia, K.N., Shin, C.T., 1982. High-Re solutions for incompressible flow using the Navier-Stokes equations and a multigrid method. Journal of Computational Physics 48, 387–411. https://doi.org/10.1016/0021-9991(82)90058-4

Guermond, J.L., Minev, P., Shen, J., 2006. An overview of projection methods for incompressible flows. Computer Methods in Applied Mechanics and Engineering 195, 6011–6045. https://doi.org/10.1016/j.cma.2005.10.010

Gueyffier, D., Li, J., Nadim, A., Scardovelli, R., Zaleski, S., 1999. Volume-of-Fluid Interface Tracking with Smoothed Surface Stress Methods for Three-Dimensional Flows. Journal of Computational Physics 152, 423–456. https://doi.org/10.1006/jcph.1998.6168

Guo, Z., 2021. Well-balanced lattice Boltzmann model for two-phase systems. Physics of Fluids 33, 031709. https://doi.org/10.1063/5.0041446

Harlow, F.H., Welch, J.E., 1965. Numerical Calculation of Time-Dependent Viscous Incompressible Flow of Fluid with Free Surface. Phys. Fluids 8, 2182. https://doi.org/10.1063/1.1761178

He, X., Chen, S., Zhang, R., 1999. A Lattice Boltzmann Scheme for Incompressible Multiphase Flow and Its Application in Simulation of Rayleigh–Taylor Instability. Journal of Computational Physics 152, 642–663. https://doi.org/10.1006/jcph.1999.6257

Hirt, C.W., Nichols, B.D., 1981. Volume of fluid (VOF) method for the dynamics of free boundaries. Journal of Computational Physics 39, 201–225. https://doi.org/10.1016/0021-9991(81)90145-5

Huang, Z., Lin, G., Ardekani, A.M., 2019. Mixed upwind/central WENO. Journal of Computational Physics 387, 455–480. https://doi.org/10.1016/j.jcp.2019.02.043

Hysing, S., Turek, S., Kuzmin, D., Parolini, N., Burman, E., Ganesan, S., Tobiska, L., 2009. Quantitative benchmark computations of two-dimensional bubble dynamics. Int. J. Numer. Meth. Fluids 60, 1259–1288. https://doi.org/10.1002/fld.1934

Inguva, V., Kenig, E.Y., Perot, J.B., 2022. A front-tracking method for two-phase flow simulation with no spurious currents. Journal of Computational Physics 456, 111006. https://doi.org/10.1016/j.jcp.2022.111006

Issa, R.I., 1986. Solution of the implicitly discretised fluid flow equations by operator-splitting. Journal of Computational Physics 62, 40–65. https://doi.org/10.1016/0021-9991(86)90099-9

Jacqmin, D., 1999. Calculation of Two-Phase Navier–Stokes Flows Using Phase-Field Modeling. Journal of Computational Physics 155, 96–127. https://doi.org/10.1006/jcph.1999.6332

Jiang, G.-S., Shu, C.-W., 1996. Efficient Implementation of Weighted ENO Schemes*. JOURNAL OF COMPUTATIONAL PHYSICS 126, 202–228.

Kang, M., Fedkiw, R.P., Liu, X.-D., n.d. A Boundary Condition Capturing Method for Multiphase Incompressible Flow 38.

Kawaguchi, Y., Tao, W.-Q., Ozoe, H., 2002. CHECKERBOARD PRESSURE PREDICTIONS DUE TO THE UNDERRELAXATION FACTOR AND TIME STEP SIZE FOR A NONSTAGGERED GRID WITH MOMENTUM INTERPOLATION METHOD. Numerical Heat Transfer, Part B: Fundamentals 41, 85–94. https://doi.org/10.1080/104077902753385027

Klostermann, J., Schaake, K., Schwarze, R., 2013. Numerical simulation of a single rising bubble by VOF with surface compression: NUMERICAL SIMULATION OF A SINGLE RISING BUBBLE BY VOF WITH SURFACE COMPRESSION. Int. J. Numer. Meth. Fluids 71, 960–982. https://doi.org/10.1002/fld.3692

Liu, H., Valocchi, A.J., Zhang, Y., Kang, Q., 2014. Lattice Boltzmann phase-field modeling of thermocapillary flows in a confined microchannel. Journal of Computational Physics 256, 334–356. https://doi.org/10.1016/j.jcp.2013.08.054

Lou, M., Liang, W., 2020. Effect of multiphase internal flows considering hydrate phase transitions on the streamwise oscillation of marine risers. Ocean Engineering 197, 106905. https://doi.org/10.1016/j.oceaneng.2019.106905

Magnini, M., Pulvirenti, B., Thome, J.R., 2016. Characterization of the velocity fields generated by flow initialization in the CFD simulation of multiphase flows. Applied Mathematical Modelling 40, 6811–6830. https://doi.org/10.1016/j.apm.2016.02.023

Majumdar, S., 1988. ROLE OF UNDERRELAXATION IN MOMENTUM INTERPOLATION FOR CALCULATION OF FLOW WITH NONSTAGGERED GRIDS. Numerical Heat Transfer 13, 125–132. https://doi.org/10.1080/10407788808913607

Mirjalili, S., Mani, A., 2021. Consistent, energy-conserving momentum transport for simulations of two-phase flows using the phase field equations. Journal of Computational Physics 426, 109918. https://doi.org/10.1016/j.jcp.2020.109918

Osher, S., Sethian, J.A., 1988. Fronts propagating with curvature-dependent speed: Algorithms based on Hamilton-Jacobi formulations. Journal of Computational Physics 79, 12–49. https://doi.org/10.1016/0021-9991(88)90002-2

Pascau, A., 2011. Cell face velocity alternatives in a structured colocated grid for the unsteady Navier-Stokes equations. Int. J. Numer. Meth. Fluids 65, 812–833. https://doi.org/10.1002/fld.2215

Patankar, S.V., Spalding, D.B., 1972. A calculation procedure for heat, mass and momentum transfer in three-dimensional parabolic flows. International Journal of Heat and Mass Transfer 15, 1787–1806. https://doi.org/10.1016/0017-9310(72)90054-3

Peskin, C.S., 1972. Flow patterns around heart valves: A numerical method. Journal of Computational Physics 10, 252–271. https://doi.org/10.1016/0021-9991(72)90065-4

Poblador-Ibanez, J., Sirignano, W.A., 2022. A volume-of-fluid method for variable-density, two-phase flows at supercritical pressure. Physics of Fluids 34, 053321. https://doi.org/10.1063/5.0086153

Ren, Y.-X., Liu, M., Zhang, H., 2007. Implementation of the Divergence-Free and Pressure-Oscillation-Free Projection Method for Solving the Incompressible Navier-Stokes Equations on the Collocated Grids. Commun. Comput. Phys. 14.

Rhie, C.M., Chow, W.L., 1983. Numerical study of the turbulent flow past an airfoil with trailing edge separation. AIAA Journal 21, 1525–1532. https://doi.org/10.2514/3.8284

Rider, W.J., Kothe, D.B., 1998. Reconstructing Volume Tracking. Journal of Computational Physics 141, 112–152. https://doi.org/10.1006/jcph.1998.5906

Sharma, D., Coquerelle, M., Erriguible, A., Amiroudine, S., 2021. Adaptive interface thickness based mobility—Phase-field method for incompressible fluids. International Journal of Multiphase Flow 142. https://doi.org/10.1016/j.ijmultiphaseflow.2021.103687

Shen, J., Yang, X., 2010. A Phase-Field Model and Its Numerical Approximation for Two-Phase Incompressible Flows with Different Densities and Viscosities. SIAM J. Sci. Comput. 32, 1159–1179. https://doi.org/10.1137/09075860X

Shu, C.-W., Osher, S., 1988. Efficient implementation of essentially non-oscillatory shock-capturing schemes. Journal of Computational Physics 77, 439–471. https://doi.org/10.1016/0021-9991(88)90177-5

Soligo, G., Roccon, A., Soldati, A., 2019. Coalescence of surfactant-laden drops by Phase Field Method. Journal of Computational Physics 376, 1292–1311. https://doi.org/10.1016/j.jcp.2018.10.021

Sui, Y., Ding, H., Spelt, P.D.M., 2014. Numerical Simulations of Flows with Moving Contact Lines. Annu. Rev. Fluid Mech. 46, 97–119. https://doi.org/10.1146/annurev-fluid-010313-141338

Sussman, M., Almgren, A.S., Bell, J.B., Colella, P., Howell, L.H., Welcome, M.L., 1999. An Adaptive Level Set Approach for Incompressible Two-Phase Flows. Journal of Computational Physics 148, 81–124. https://doi.org/10.1006/jcph.1998.6106

Wang, C., Wu, Q., Huang, B., Wang, G., 2018. Numerical investigation of cavitation vortex dynamics in unsteady cavitating flow with shock wave propagation. Ocean Engineering 156, 424–434. https://doi.org/10.1016/j.oceaneng.2018.03.011

Wu, W.B., Liu, Y.L., Zhang, A.M., 2017. Numerical investigation of 3D bubble growth and detachment. Ocean Engineering 138, 86–104. https://doi.org/10.1016/j.oceaneng.2017.04.023

Xia, Q., Kim, J., Li, Y., 2022. Modeling and simulation of multi-component immiscible flows based on a modified Cahn–Hilliard equation. European Journal of Mechanics - B/Fluids 95, 194–204. https://doi.org/10.1016/j.euromechflu.2022.04.013

Xiao, Y., Zeng, Z., Zhang, L., Wang, J., Wang, Y., Liu, H., Huang, C., 2022a. A spectral element-based phase field method for incompressible two-phase flows. Physics of Fluids 34, 022114. https://doi.org/10.1063/5.0077372

Xiao, Y., Zeng, Z., Zhang, L., Wang, J., Wang, Y., Liu, H., Huang, C., 2022b. A highly accurate bound-preserving phase field method for incompressible two-phase flows. Physics of Fluids 5.0103277. https://doi.org/10.1063/5.0103277

Ying, W., Henriquez, C.S., 2007. A kernel-free boundary integral method for elliptic boundary value problems. Journal of Computational Physics 227, 1046–1074. https://doi.org/10.1016/j.jcp.2007.08.021

Yuan, H.-Z., Shu, C., Wang, Y., Shu, S., 2018. A simple mass-conserved level set method for simulation of multiphase flows. Physics of Fluids 30, 040908. https://doi.org/10.1063/1.5010152

Zhang, L., Zeng, Z., Xie, H., Zhang, Y., Lu, Y., Yoshikawa, A., Mizuseki, H., Kawazoe, Y., 2014. A comparative study of lattice Boltzmann models for incompressible flow. Computers & Mathematics with Applications 68, 1446–1466. https://doi.org/10.1016/j.camwa.2014.09.010

Zhang, T., Wu, J., Lin, X., 2019. An interface-compressed diffuse interface method and its application for multiphase flows. Physics of Fluids 31, 122102. https://doi.org/10.1063/1.5116035

Zheng, H.W., Shu, C., Chew, Y.T., 2006. A lattice Boltzmann model for multiphase flows with large density ratio. Journal of Computational Physics 218, 353–371. https://doi.org/10.1016/j.jcp.2006.02.015